\documentclass[pdflatex,sn-mathphys-ay]{sn-jnl}

\usepackage{graphicx}%
\usepackage{multirow}%
\usepackage{amsmath,amssymb,amsfonts}%
\usepackage{amsthm}%
\usepackage{mathrsfs}%
\usepackage[title]{appendix}%
\usepackage{xcolor}%
\usepackage{textcomp}%
\usepackage{manyfoot}%
\usepackage{booktabs}%
\usepackage{algorithm}%
\usepackage{algorithmicx}%
\usepackage{algpseudocode}%
\usepackage{listings}%
\usepackage{graphicx}
\usepackage{subcaption}

\theoremstyle{thmstyleone}%
\theoremstyle{thmstyletwo}%

\theoremstyle{thmstylethree}%

\begin{document}
\title{The Impact of COVID-19 on Twitter Ego Networks: Structure, Sentiment, and Topics} 


\author*[1,2]{\fnm{Kamer} \sur{Cekini}}\email{kamer.cekini@phd.unipi.it}

\author[1]{\fnm{Elisabetta} \sur{Biondi}}\email{elisabetta.biondi@iit.cnr.it}

\author[1]{\fnm{Chiara} \sur{Boldrini}}\email{chiara.boldrini@iit.cnr.it}

\author[1]{\fnm{Andrea} \sur{Passarella}}\email{andrea.passarella@iit.cnr.it}

\author[1]{\fnm{Marco} \sur{Conti}}\email{marco.conti@iit.cnr.it}


\affil[1]{\orgname{IIT-CNR}, \orgaddress{\street{Via G. Moruzzi, 1}, \city{Pisa}, \postcode{56124}, \country{Italy}}}

\affil[2]{\orgname{Universit\`a di Pisa}, \orgaddress{\street{Largo B. Pontecorvo, 3}, \city{Pisa}, \postcode{56127}, \country{Italy}}} 

\abstract{Lockdown measures, implemented by governments during the initial phases of the COVID-19 pandemic to reduce physical contact and limit viral spread, imposed significant restrictions on in-person social interactions. Consequently, individuals turned to online social platforms to maintain connections. Ego networks, which model the organization of personal relationships according to human cognitive constraints on managing meaningful interactions, provide a framework for analyzing such dynamics. The disruption of physical contact and the predominant shift of social life online potentially altered the allocation of cognitive resources dedicated to managing these digital relationships. This research aims to investigate the impact of lockdown measures on the characteristics of online ego networks, presumably resulting from this reallocation of cognitive resources. To this end, a large dataset of Twitter users was examined, covering a seven-year period of activity. Analyzing a seven-year Twitter dataset—including five years pre-pandemic and two years post—we observe clear, though temporary, changes. During lockdown, ego networks expanded, social circles became more structured, and relationships intensified. Simultaneously, negative interactions increased, and users engaged with a broader range of topics, indicating greater thematic diversity. Once restrictions were lifted, these structural, emotional, and thematic shifts largely reverted to pre-pandemic norms—suggesting a temporary adaptation to an extraordinary social context.}




\maketitle

\section{Introduction}
\label{sec:introduction}

The COVID-19 pandemic, along with the resulting lockdowns, brought about significant societal transformations worldwide. With widespread stay-at-home mandates, people were suddenly cut off from in-person interactions, leading to a rapid and massive migration of social exchanges into digital environments. This abrupt transition led to an unprecedented surge in activity across all major social media platforms, fueled by the constraints of physical distancing~\citep{ford_rojas_coronavirus_2020,schultz_keeping_2020}.

While user behavior on social media—and particularly on Twitter—during the pandemic has been the focus of considerable research~\citep{huang2020twitter,mattei2021italian,miyazaki2023fake}, one aspect remains underexplored: how lockdowns specifically reshaped the fine-grained fabric of social interaction within these platforms. In this work, we seek to fill this gap by analyzing the transformation of online social dynamics using the framework of ego networks. We focus not only on how the structural aspects of these networks evolved (such as their size and the number of social circles), but also on the changes in relationship characteristics (e.g., positive or negative sentiment) and the thematic content of user interactions.

\begin{figure}[!b]
    \centering
    \includegraphics[scale=0.15]{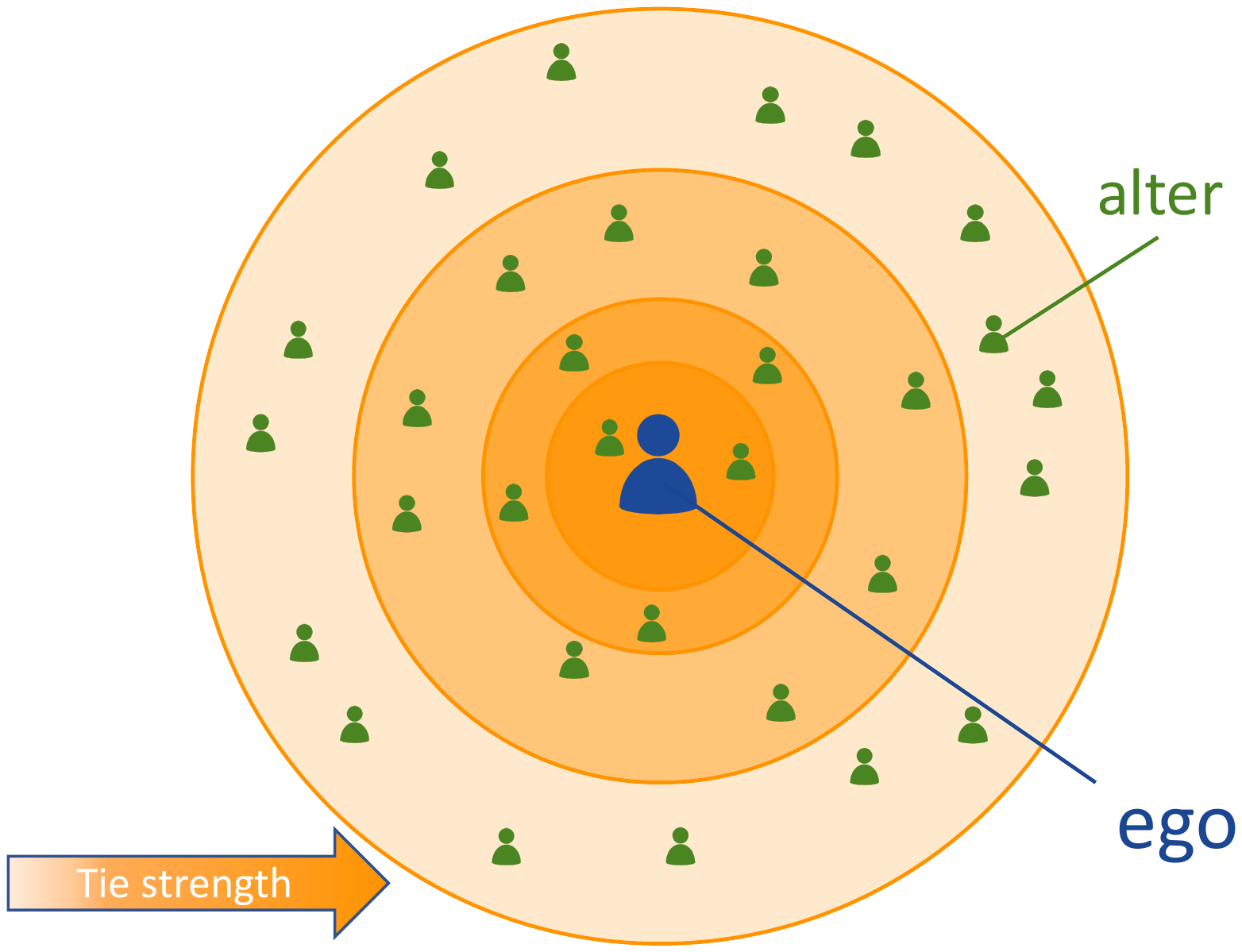}
    \caption{The ego network model.}
    \label{fig:egonet}
\end{figure}

Ego networks, rooted in evolutionary anthropology~\citep{Dunbar_1995}, revolve around a central individual (the ego) and their direct social contacts, known as alters. These networks are typically represented as a series of concentric layers, with the most intimate and strongest relationships residing in the innermost circle and weaker, more distant ties occupying the outer layers (Figure~\ref{fig:egonet}). This layered configuration mirrors the gradation of closeness found in human social relationships. Commonly, ego networks are composed of approximately 5, 15, 50, and 150 alters~\citep{zhou_sornette_hill_dunbar_2005}, with each circle being roughly three times the size of the previous one~\citep{hill_dunbar_2003}. Importantly, these networks include only those relationships that the ego actively maintains—ties that are meaningful and sustained over time. Connections that are no longer actively engaged are often referred to as the inactive portion of the ego network.

In this work, we address key research questions regarding the impact of lockdowns on user ego networks. To what extent did their structure change? How did the polarity of relationships and the thematic content of interactions evolve during this period? Finally, were these changes lasting, or did these network characteristics return to their original levels after lockdowns? For our analysis we use a dataset comprising 1286 Twitter\footnote{Since our dataset was collected before Twitter changed its name to X, in this work we refer to the platform with its former name.} users, and we study how the discussed properties of ego networks of these users change over the years. We already presented the results about the structural properties of ego networks in~\citep{cekini2024socialisolationdigitalconnection}. In this work, we extend that analysis by including an investigation on the polarity and thematic diversity of social relationships in Twitter ego networks. Together, the current work and \citep{cekini2024socialisolationdigitalconnection} provide a comprehensive overview of the impact of COVID-19 lockdowns on the cognitive engagement of users with online social networks.
Our main findings are summarized below, covering both \cite{cekini2024socialisolationdigitalconnection} and the current work:
\begin{itemize}
    \item \textbf{Ego networks expanded during lockdown}, with more alters and additional circles, especially in the outer layers.
    \item \textbf{Alters tended to move toward inner circles}, suggesting strengthened or more intimate online relationships.
    \item \textbf{Negative interactions within ego networks increased}, reflecting the emotional strain of the lockdown period, consistent with findings in the literature.
    \item \textbf{Users engaged with a broader range of topics}, showing increased thematic diversity in their online communication.
    \item \textbf{These changes were mostly temporary adaptations}, with metrics returning to pre-pandemic levels once restrictions were lifted.
\end{itemize}
In short, our results indicate that users’ cognitive engagement online intensified during the lockdown, both in terms of social interactions and thematic exploration. The reduction in offline social opportunities freed up cognitive resources, enabling users to invest more deeply in their online networks. Once restrictions were lifted, attention shifted back to offline life, leading to a return to pre-pandemic patterns of online behavior.

\section{Background and Related Work}
\label{sec:related_work}

As discussed in Section~\ref{sec:introduction}, \emph{ego networks} capture the relationships between a central individual—the \emph{ego}—and their social contacts, or \emph{alters}. These networks take the form of localized subgraphs, where the strength of each connection (often measured through metrics like interaction frequency~\citep{hill_dunbar_2003,pollet_roberts_dunbar_2011,arnaboldi2013ego} reflects the closeness of the relationship.
Organizing these connections by strength gives rise to what are known as \emph{intimacy layers} (Figure~\ref{fig:egonet}). These layers are typically composed of approximately 5, 15, 50, and 150 individuals, with emotional proximity decreasing as one moves outward. The outermost layer—encompassing roughly 150 alters—is commonly referred to as \emph{Dunbar’s number}, which denotes the cognitive limit for the number of stable social relationships a person can maintain~\citep{hill_dunbar_2003,zhou_sornette_hill_dunbar_2005}.
This stratified pattern is thought to emerge from cognitive limitations described by the \emph{social brain hypothesis}~\citep{dunbar_1998}, which posits that brain capacity constrains the number of social ties we can sustain. As a result, individuals optimize their social effort across these layers~\citep{sutcliffe2012relationships}. One of the most consistent features of ego networks is the \emph{scaling ratio} between successive layers—usually close to three—which has been observed in both offline and online settings~\citep{Dunbar2015,zhou_sornette_hill_dunbar_2005}.
\emph{Dunbar’s structure} has been validated across multiple forms of real-world communication~\citep{roberts2009,hill_dunbar_2003,miritello2013}, as well as in \emph{online social networks} (OSNs)~\citep{Dunbar2015}, supporting the idea that digital social behavior is still shaped by the same cognitive boundaries found in offline interactions.
Subsequent studies have delved into the mechanisms behind \emph{tie strength} and \emph{ego network formation}~\citep{gonccalves2011modeling,quercia2012social}, their implications for the spread and \emph{diversity of information}~\citep{aral2011diversity}, the interplay between ego networks and \emph{sentiment in relationships}~\citep{tacchi2024keep}, and how these structures can be leveraged for tasks like \emph{link prediction}~\citep{toprak2022harnessing}. 

Complementing structural analyses, research has also examined the polarity (positive vs. negative nature) of relationships within ego networks \citep{tacchi2024keep}. Studies on platforms like Twitter suggest that while online ego networks share structural properties with offline counterparts, they may exhibit a higher prevalence of negative ties, particularly in inner social circles \citep{tacchi2024keep}. This phenomenon might be linked to platform dynamics that potentially amplify conflict or increase the visibility of negative interactions \citep{Ferrara_2015}.

Alongside investigations into structure and polarity, research also explores the thematic content of interactions within ego networks. Studies indicate that discussion topics significantly influence network characteristics. For example, \citet{tacchi2024jointeffectculturediscussion} demonstrated that engagement with specific or polarizing subjects is associated with higher relationship negativity on Twitter. To analyze this thematic dimension, researchers utilize topic modeling techniques; for instance, the BERTopic framework has been employed in relevant studies \citep{tacchi2024jointeffectculturediscussion, Ollivier_2022} to identify discussion themes within Twitter data.

%
%

\section{The Dataset} 
\label{sec:the_dataset}

Our goal is to analyze how the COVID-19 pandemic and related lockdown measures affected the structure, polarity, and topics discussed in online social networks. We use March 1, 2020, as the reference point for the onset of widespread lockdowns, particularly in Europe and North America~\citep{wikipediaCOVID19}. We consider a 7-year window spanning from March 1, 2015, to March 1, 2022, to observe user behavior before, during, and after the pandemic. To this aim, we used the dataset collected by~\cite{cekini2024socialisolationdigitalconnection}. The collection started from user Roberto Burioni, a prominent Italian virologist active against anti-vaccination campaigns, and proceeded in a snowball fashion. The dataset was cleaned to remove bots. Similar to~\cite{cekini2024socialisolationdigitalconnection},  we define yearly intervals $I_k = [\text{March 1, 2015} + k \text{ years}, \text{March 1, 2016} + k \text{ years}]$ for $k = 0, \ldots, 6$. The lockdown marks the boundary between  $I_4$ and $I_5$. We retain users who are regular and active in all intervals, i.e., 
$$
 R_0 \cap R_1 \cap R_2 \cap R_3 \cap R_4 \cap R_5 \cap R_6.
$$
A user is considered regular if they engage in interactions (mentions, replies, retweets) in at least 50\% of the months in a given period, and active if the time since their last tweet is not significantly longer than their typical intertweet interval (specifically, less than 6 months longer). These are standard definitions used in the related literature~\citep{arnaboldi2013ego,arnaboldi2017structure,boldrini2018twitter}.
This yields a core group of 1,627 highly social users, enabling consistent longitudinal analysis of ego network evolution and lockdown effects. While not fully representative, this group offers valuable insights into broader social trends. 

As shown in Figure~\ref{fig:egonet_size_all} in Appendix~\ref{app:dataset_context}, some users had extremely large active ego networks (over 500), skewing the data. We treated these as outliers using the interquartile range (IQR) method, applied separately to each period $I_k$.

\begin{figure}[t!]
    \centering
    \includegraphics[width=\textwidth]{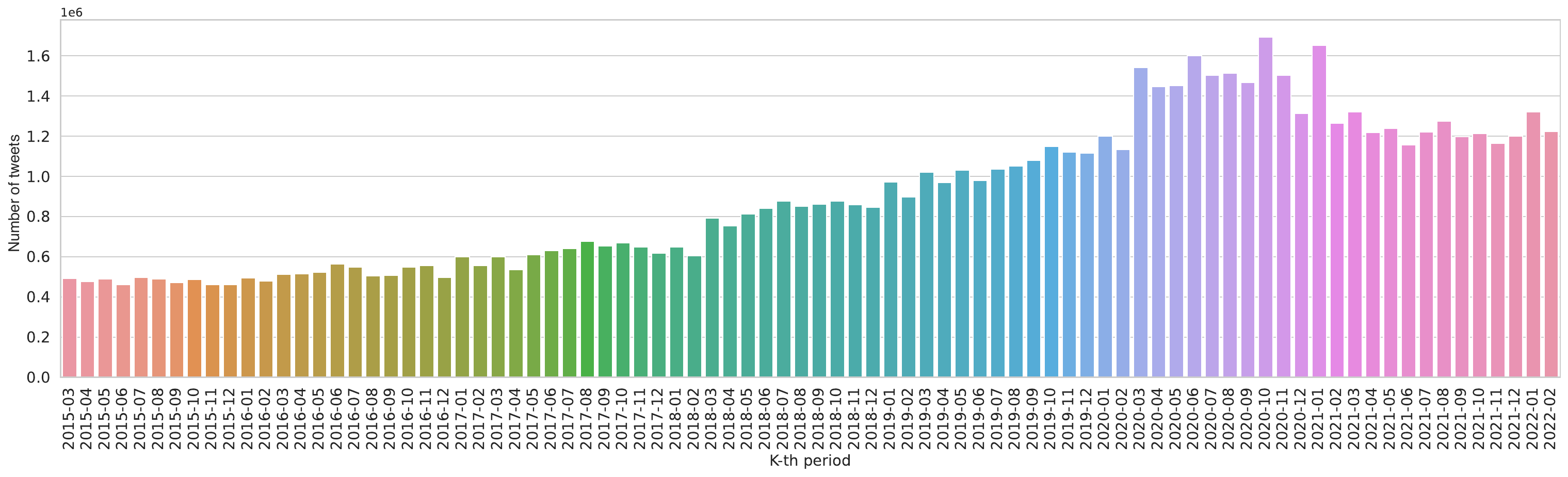}
    \caption{Total number of tweets considered}
    \label{fig:number_tweets}
    \vspace{-10pt} 
\end{figure}

\begin{figure}[h!] 
\centering 
\begin{minipage}[t]{0.45\textwidth}
    \centering 
    \includegraphics[width=0.8\linewidth]{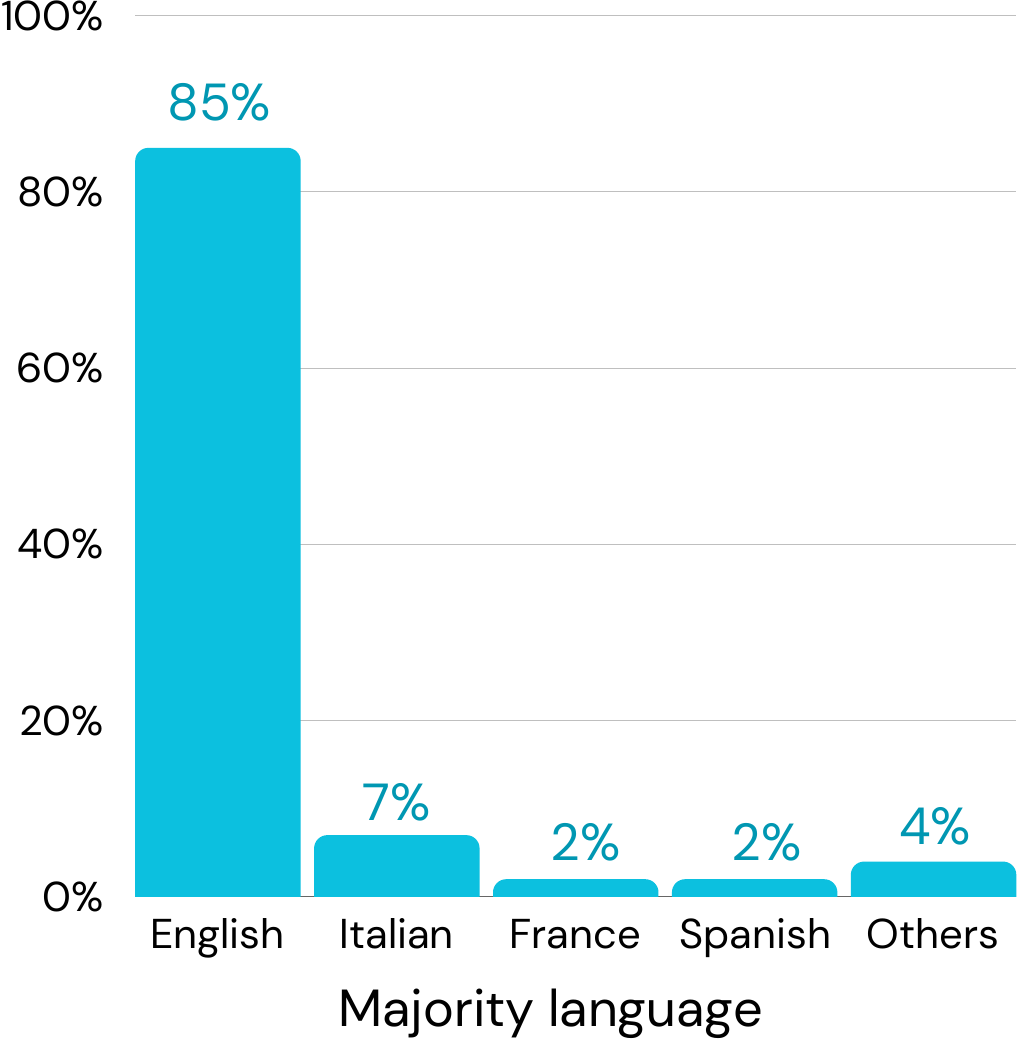}
    \caption{Percentage distribution of majority languages among the 1,286 users analyzed.}
    \label{fig:majority_language}
\end{minipage}
\hfill 
\begin{minipage}[t]{0.45\textwidth}
    \centering 
    \includegraphics[width=0.8\linewidth]{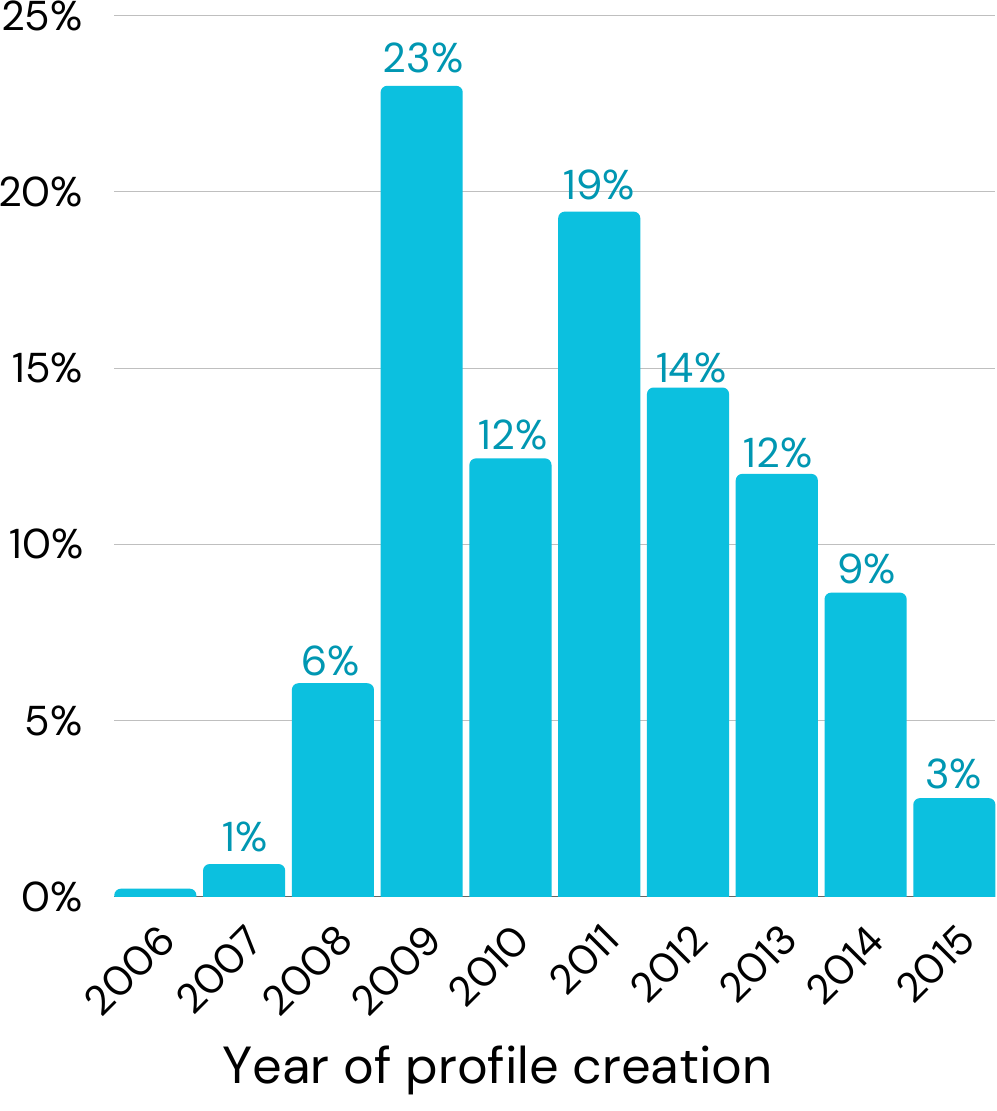}
    \caption{Profile creation dates of the 1,286 analyzed users.}
    \label{fig:year_profile_creation}
\end{minipage}
\vspace{-10pt}
\end{figure}

We excluded users identified as outliers in any interval ($\bigcap_{i=0}^6 R_{I_i} \setminus \bigcup_{i=0}^6 O_{I_i},$ where $O_{I_i}$ contains outliers based on ego network size during $I_i$. This reduced the dataset to 1,286 users. Table~\ref{tab:users_filtered} in Appendix~\ref{app:dataset_context} provides a summary of all the filters applied. These 1,286 users collectively generated over 67 million tweets. The distribution of tweet activity over the observed time window, showcasing a significant peak in March 2020 that aligns with the onset of the lockdown, is illustrated in Figure~\ref{fig:number_tweets}. English is the predominant language among the users analyzed, with 87\% tweeting in English. Following that, 7\% of users write in Italian, while the remaining users communicate in French, Spanish, and other languages (see Figure~\ref{fig:majority_language} for the complete distribution). 
Additionally, the analysis of profile creation dates reveals that the majority of the 1,286 users created their profiles between 2009 and 2013, as shown in Figure~\ref{fig:year_profile_creation}.

\section{Methodology} \label{sec:methodology}
This section details the methodologies employed to analyze the impact of COVID-19 lockdowns on ego networks across multiple dimensions, as outlined in Section~\ref{sec:introduction}. We first describe the procedures for extracting the fundamental \textbf{structural properties} of ego networks (Section~\ref{sec:ego_network_extraction}). Subsequently, we outline the method used to determine \textbf{relationship polarity} (positive/negative), enabling the analysis of signed ego networks (Section~\ref{sec:signed_ego_net_extraction}). We finally detail the techniques used for \textbf{semantic analysis}, specifically topic modeling, to understand the content of interactions (Section~\ref{sec:semantic_ego_network}). 

\subsection{Extracting Ego Networks}
\label{sec:ego_network_extraction}

This section provides a concise overview of the process used to build ego networks.
The initial step involves determining the contact frequency between the ego and its alters. For online social ties, this frequency is defined as the number of direct interactions  (specifically replies, mentions, and retweets) divided by the duration of the relationship in years. Consequently, the interaction frequency, which serves as an indicator of social closeness, between an ego $u$ and an alter $j$ during the time interval $I_k$ is computed using the following expression:
\begin{equation}
w_{uj}^{(i)} = \frac{n_{reply}^{(u,j)} + n_{mention}^{(u,j)} + n_{retweet}^{(u,j)}}{I_i},
\end{equation}
where $n_*$ is the number of interactions of type * (one of replies, mentions, and retweets) from ego $u$ to alter $j$ and $I_i$ is the time length of the $i$-th period considered in our analysis. All the relationships with contact frequency $w_{uj}^{(i)} \geq 1$ are called \emph{active} and are part of the \emph{active} ego network of user $u$. 
After calculating the intimacy of the relationships as mentioned in the previous paragraph, we can group the active relationships into intimacy levels. To this aim, and similarly to the related literature \citep{boldrini2018twitter,toprak2022harnessing,tacchi2024keep}, we use the Mean Shift algorithm. The advantage of Mean Shift, against, e.g., more traditional clustering methods like $k$-means, is that it automatically selects the optimal number of clusters. Each of the clusters found by Mean Shift corresponds to a \emph{ring} $\mathcal{R}$ in the ego network (Figure~\ref{fig:egonet}), with $\mathcal{R}_1$ being the one with the highest average contact frequency (i.e., intimacy). Then, circles $\mathcal{C}$ are obtained as the union set of concentric rings. Thus, it holds that $\mathcal{C}_{k} = \mathcal{C}_{k-1} \cup \mathcal{R}_k$, with initial condition $\mathcal{C}_1 = \mathcal{R}_1$. The active ego network size is thus the size of the largest circle. Note that we compute ego networks (hence their circles and rings) for each period $I_i$, so we will have circles $\mathcal{C}_{k}^{(i)}$ and rings $\mathcal{R}_k^{(i)}$.

\subsection{Extracting the Polarity (Signs) of Relationships}
\label{sec:signed_ego_net_extraction}

Unlike traditional ego networks, signed ego networks incorporate information about the polarity of relationships between nodes. In this framework, each connection between the ego $i$ and an alter $j$ is labelled as either positive or negative. Positive links denote relationships marked by trust, affinity, and cooperative behavior, reflecting a high degree of social cohesion~\citep{Maniu2011}. In contrast, negative links capture antagonistic or distrustful relationships, signaling disagreement, conflict, or social distance between the individuals involved~\citep{tacchi2024keep}.

To determine the polarity of the relationship between an ego and an alter, we begin by classifying their \textit{individual interactions} as positive, negative, or neutral. Here, “interactions” include replies, mentions, quotes, and retweets. Following~\cite{tacchi2024keep}, retweets are considered neutral, as they reflect content resharing rather than direct engagement. The remaining interaction types are analyzed using the BertTweet model—a variant of the RoBERTa (Robustly Optimized BERT Pretraining Approach) architecture tailored for Twitter data.
This model has been fine-tuned on an extensive corpus of over 845 million English-language tweets collected between January 2012 and August 2019, along with an additional 5 million COVID-19-related tweets. As a result, BertTweet is well-equipped to handle Twitter’s distinctive linguistic features, such as abbreviations, slang, emojis, and platform-specific symbols\citep{huang2020twitter}.

After classifying individual interactions, we determine the overall polarity of the relationship between two users by analyzing the distribution of their interactions. Specifically, we compute the proportion of positive and negative interactions between each pair of nodes. To ensure consistency and comparability with recent studies on signed networks on Twitter, we adopt the methodology proposed by~\cite{tacchi2024keep}: a relationship is classified as negative if more than 17\% of the interactions are negative.
This 17\% threshold is grounded in psychological research, which indicates that a relatively small proportion of negative exchanges can signal instability in a relationship, while a lower proportion is typically associated with stable, positive ties~\citep{tacchi2024keep, Gottman1995, HartRisley1995}.
Although some interactions may be neutral (e.g., retweets), each relationship is ultimately labeled as either positive or negative, with no neutral category. This binary classification reflects the assumption that the effort invested in maintaining communication correlates with relational strength; thus, in the absence of explicit negativity, interactions are presumed to be positive~\citep{tacchi2024keep}.
While \cite{tacchi2024keep} characterized properties of signed Twitter networks, our contribution focuses distinctively on their temporal evolution, examining shifts in relationship polarity in response to the COVID-19 lockdown.

\subsection{Extracting the Semantic Ego Network}
\label{sec:semantic_ego_network}

Beyond the structural dynamics (such as network size and number of circles, analyzed in Section~\ref{sec:ego_network_extraction}) and the nature of relationships (positive or negative, Section~\ref{sec:signed_ego_net_extraction}), we analyze the content of interactions. This offers an additional dimension for understanding the lockdown's impact on online social behavior, potentially revealing shifts in the topics discussed and interests expressed by users during this period of profound social change.

In this work, we define a user's \textit{semantic ego network} in a given period as the set of \textit{topics} extracted from the textual content of their \textit{social interactions}. The objective is to determine whether the lockdown influenced not only the structure and nature (positive/negative) of online interactions but also their thematic content.
To achieve this goal, we employ topic modeling techniques applied to the textual content generated by users in our dataset. The following subsections detail the text data preparation process (Section~\ref{sec:pre_processing}), the chosen topic modeling approach based on BERTopic (Section~\ref{sec:BERTopic}).

\subsubsection{Text Pre-Processing}
\label{sec:pre_processing}

The topic analysis is based on the dataset described in Section~\ref{sec:the_dataset}, which includes users defined as $\bigcap_{i=0}^{6} R_{I_i} \setminus \bigcup_{i=0}^{6} O_{I_i}$ and their respective social tweets (i.e., replies, quotes, mentions and retweets).
For each time interval $I_i$ and for each user, referred to as ego $j$, we define $Tw_{I_i}^j$ as the set of all social tweets published during the period $I_i$ by the ego $j$, and $Re_{I_i}^j \subseteq Tw_{I_i}^j$ as the subset of retweets made by the same ego during the same period.
Since topic extraction requires the full text of tweets, and the dataset does not provide the textual content that retweets refer to, these are excluded from the analysis. Therefore, the set of tweets analyzed is defined as $Ts = \bigcup_{i,j} (Tw_{I_i}^j \setminus Re_{I_i}^j)$.

A pre-processing procedure is applied to the dataset $Ts$ to improve the quality of the text. In the first stage, links and references to images and videos are removed, as multimedia content is not accessible through the text field alone. Subsequently, to avoid distortions in the topic representation caused by repetitive messages or spam, duplicates are removed. The final dataset (free from retweets, duplicates, links, and multimedia content), denoted as $Ts'$, consists of $N = 4,850,558$ textual entries.
This pre-processing ensures that the data used in the analysis are as representative as possible of social communication, free from elements that might introduce noise or distortions.

\subsubsection{Topic Modeling with BERTopic}
\label{sec:BERTopic}

BERTopic~\citep{grootendorst2022bertopicneuraltopicmodeling} is a topic modeling framework particularly well-suited for analyzing tweets, which are typically short, informal, and often noisy. It uses sentence embeddings to capture the meaning of text in context, allowing it to go beyond simple keyword matching.
%
%
%
The BERTopic process is divided into four phases: (i) extraction of text embeddings, (ii) dimensionality reduction~\citep{mcinnes2020umapuniformmanifoldapproximation}, (iii) clustering of documents using HDBSCAN~\citep{Campello2013HDBSCAN}, and, finally, (iv) the representation of topics through c-TF-IDF~\citep{grootendorst2022bertopicneuraltopicmodeling} and the calculation of topic embeddings.

Considering the presence of tweets in various languages (see the \textit{Appendix} for details), the SBERT~\citep{reimers2019sentencebertsentenceembeddingsusing} multilingual model \href{https://huggingface.co/sentence-transformers/paraphrase-multilingual-mpnet-base-v2}{paraphrase-multilingual-mpnet-base-v2} was utilized for phase (i). The obtained embeddings are high-dimensional (we used the standard 768 dimensions), so their dimensionality is reduced with UMAP~\citep{mcinnes2020umapuniformmanifoldapproximation}, to improve processing efficiency while preserving the local relationships among points~\citep{Allaoui2020UMAP} (phase (ii)). In phase (iii), the Hierarchical Density-Based Spatial Clustering (HDBSCAN)~\citep{Campello2013HDBSCAN} algorithm is applied. As an extension of DBSCAN, HDBSCAN is designed to identify clusters with varying density while automatically detecting outliers, avoiding the forced assignment of unrelated documents. Once the clusters are obtained, in phase (iv) BERTopic assigns each topic a set of representative words that best characterize that cluster, highlighting words that are frequent in one cluster but less common across others. To achieve this, BERTopic employs a variant of the classic Term Frequency-Inverse Document Frequency (TF-IDF), called class-based TF-IDF (c-TF-IDF)~\citep{grootendorst2022bertopicneuraltopicmodeling}.

The performance of the BERTopic pipeline is highly sensitive to the configuration of its hyperparameters. To optimize results, we conduct a grid search over four key parameters: \texttt{n\_components} and \texttt{n\_neighbors} (UMAP), and \texttt{min\_cluster\_size} (HDBSCAN). Full details of the search space are provided in Appendix~\ref{subsec:appendix_hyperparam}.
Traditionally, topic modeling performance has been evaluated using two complementary metrics: Topic Coherence~\citep{Bouma2009Normalized} and Topic Diversity~\citep{Dieng2020Topic}. However, these metrics have important limitations. They treat words in isolation, ignoring the contextual richness captured by semantic embeddings. As a result, semantically equivalent terms---such as ``Neural Network'' and ``N.N.'' or their multilingual variants (e.g., ``Neural Network'' and ``rete neurale'' in Italian)---are considered unrelated, leading to inaccurate evaluations, especially in multilingual settings.
Another drawback is that these metrics do not directly assess clustering quality. Instead, they infer performance based on the lexical representation of clusters. Consequently, even when the clustering itself is meaningful, a poor selection of representative words can negatively impact coherence and diversity scores.
Since our primary focus is on the quality of the cluster partitioning rather than on achieving an interpretable lexical labeling of topics, these traditional metrics are not fully aligned with our evaluation goals.

To address this, we adopt DBCV (Density-Based Clustering Validation)~\citep{moulavi2014density}, an index specifically designed to evaluate the quality of clusterings produced by density-based algorithms, such as DBSCAN~\citep{ester1996density}, OPTICS~\citep{ankerst1999optics}, and HDBSCAN~\citep{Campello2013HDBSCAN}. 
%
DBCV computes a validity score for each cluster by combining two key aspects: the internal density and the separation between clusters. The overall clustering quality is then derived as a weighted average of these scores, where each cluster’s contribution is proportional to its size. Without delving into the technical implementation details - which can be found in the original paper - the core idea of the metric is captured by the formula:
\[
DBCV(C) = \sum_{i=1}^{l} \frac{|C_i|}{|O|} V_C(C_i)
\]
where $C$ represents the set of clusters, $l$ is the total number of clusters, $O$ denotes the entire dataset. $V_C(C_i) \in [-1, 1]$ quantifies the quality of cluster $C_i$ based on two factors: internal density, which measures how uniformly the data points within the cluster are distributed, and separation, which evaluates how distinct the density of the cluster is compared to that of other clusters.
Scores approaching +1 indicate that a cluster has high internal cohesion and is well separated from others, whereas scores near -1 suggest overlapping clusters or poorly defined density regions. These properties ensure that DBCV produces an interpretable and meaningful score within the interval~$[-1, 1]$.

Calculating DBCV for the entire dataset with $N = 4,850,558$ data points is extremely time-consuming and computationally intensive. Therefore, we computed DBCV on a sampled subset of 500,000 elements - approximately 10\% of the original data. The sampling was conducted proportionally, drawing more points from larger clusters than from smaller ones to maintain the dataset's original distribution. 
The optimal configuration identified through grid search (see Appendix~\ref{subsec:appendix_hyperparam}, Table~\ref{tab:hyperpar_dbcv} for top results) is $(n\_components = 2, n\_neighbors = 10, min\_cluster\_size = 20)$. This configuration, achieving the highest sampled DBCV score (0.11766) and the lowest percentage of outliers (approximately 72.66\%), will be used to generate the topics analyzed in the remainder of the paper.
Notably, this optimal setup still classifies a significant portion of the tweets (approximately 72.66\%) as outliers or noise. Such a high outlier percentage is not unexpected when applying density-based clustering like HDBSCAN to noisy, short-text datasets like Twitter, where many individual messages may lack strong thematic coherence.

\section{Metrics Definition}
\label{sec:metrics}
To analyze the impact of the COVID-19 lockdown on user interactions, we track several key metrics across the yearly periods $I_i$ (for $i=0, \ldots, 6$) for each user $u$ (or $j$). These metrics capture structural, relational, and semantic aspects of the ego networks. 
For the \textbf{structural aspects}, we measure the active ego network size $|A_i^u|$ (i.e., the number of alters contacted at least once a year), the optimal number of circles (as found by Mean Shift), and individual circle sizes. 
Regarding \textbf{relational polarity}, we track the percentage of negative relationships ($N_i^u$) and positive relationships ($P_i^u$) within the ego network, calculated based on interaction sentiment. 
Finally, for the \textbf{semantic dimension}, we assess thematic diversity by measuring the number of unique topics $|T_{I_i}^j|$ discussed by user~$j$, derived from topic modeling. 

A core component of our analysis, presented in Section~\ref{sec:results}, 
involves examining the \textbf{temporal evolution} of these metrics, specifically comparing periods before, during ($I_5$), and after the main lockdown phase ($I_6$). To quantify changes between consecutive periods ($I_i, I_{i+1}$) for any generic user-level metric $X_i^u$ that varies over time (e.g., network size, percentage of negative ties, number of topics), we frequently utilize the concept of \textbf{growth rate}:
\begin{equation} 
\label{eq:growth_rate} 
G_{[i,i+1]}^u(X) = \frac{X^u_{i+1} - X^u_i}{X^u_i}
\end{equation}
A positive growth rate indicates an increase in quantity $X$, while a negative rate indicates a decrease. As detailed in Section~\ref{sec:results}, we often assess the statistical significance of observed temporal trends by applying tests (primarily t-tests) to the distribution of these growth rates or to the difference between consecutive growth rates across users, allowing us to rigorously evaluate the impact associated with the lockdown event.

\section{Results}
\label{sec:results}
This section presents the results of our analyses investigating the impact of the COVID-19 pandemic lockdown on Twitter users' ego networks, following the methodologies detailed in Section~\ref{sec:methodology}. We structure the presentation according to the three main dimensions explored: first, we recap the main findings on the changes in \textbf{ego network structure} (Section~\ref{sec:ego_network_analysis}); second, we analyze the shifts in \textbf{relationship polarity} using the signed ego network framework (Section~\ref{sec:res_signed_ego_network}); and third, we investigate the evolution of \textbf{thematic content} through topic modeling (Section~\ref{sec:semantic_topic_evolution}).

\subsection{Ego network Evolution During the COVID-19 Pandemic}
\label{sec:ego_network_analysis}

In this section, we summarise the main findings presented in detail in \citep{cekini2024socialisolationdigitalconnection}, in order to provide a comprehensive view of the impact of lockdown measures on Twitter users' online ego network. Our analysis focuses on the evolution of ego networks during the seven-year time window and is twofold. First, we discuss whether the decrease in in-person socializing due to lockdowns led to an increase in online activity within the network, exploring the possibility that the cognitive effort previously directed towards offline interactions shifted online. Second, we examine the dynamic movement within the ego networks, highlighting the changes occurring between different circles.

\begin{figure}[t!]
     \centering
     \subfloat[Ego network sizes]{
         \includegraphics[width=0.48\textwidth]{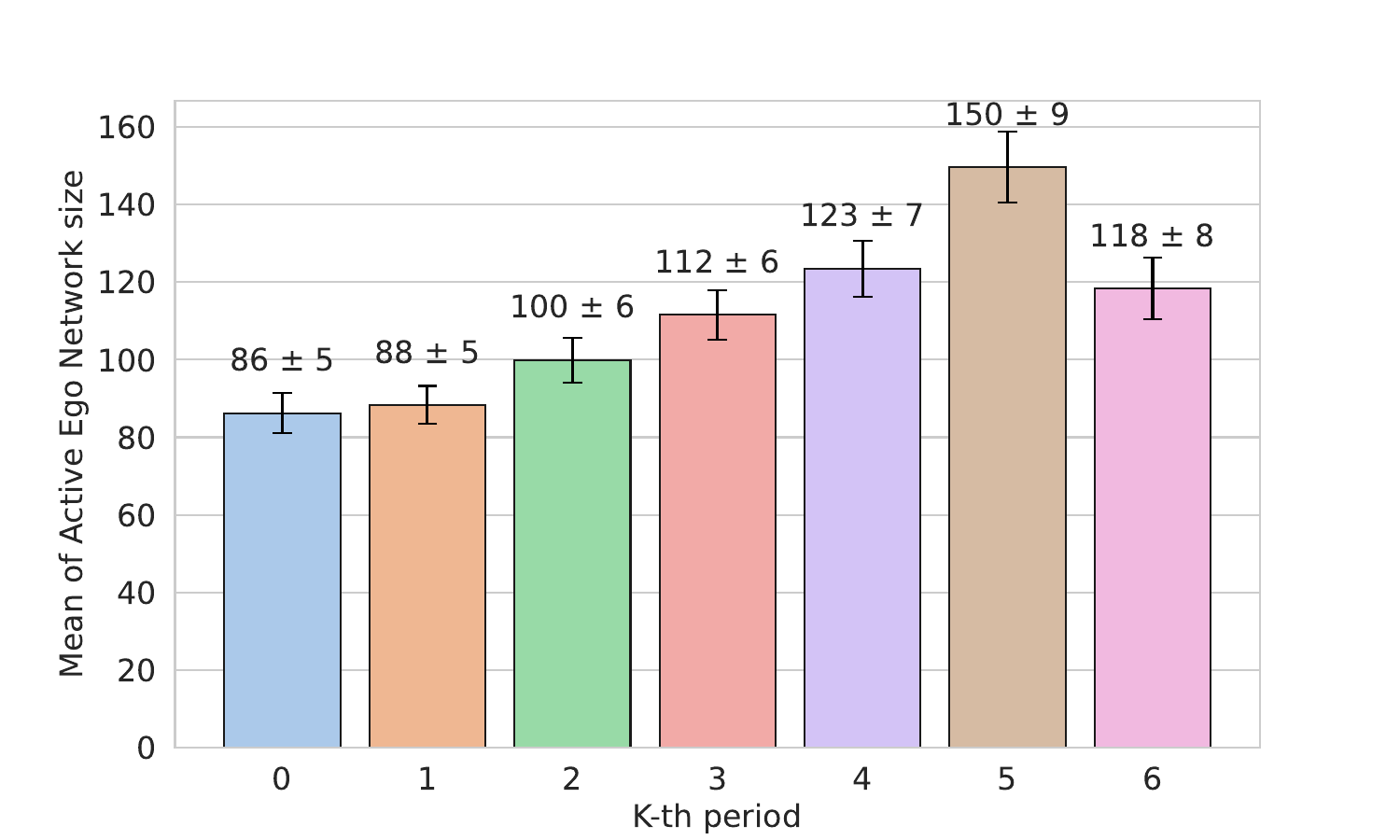}
     }\hfill
     \subfloat[Growth rates of difference of ego network sizes]{
         \includegraphics[width=0.48\textwidth]{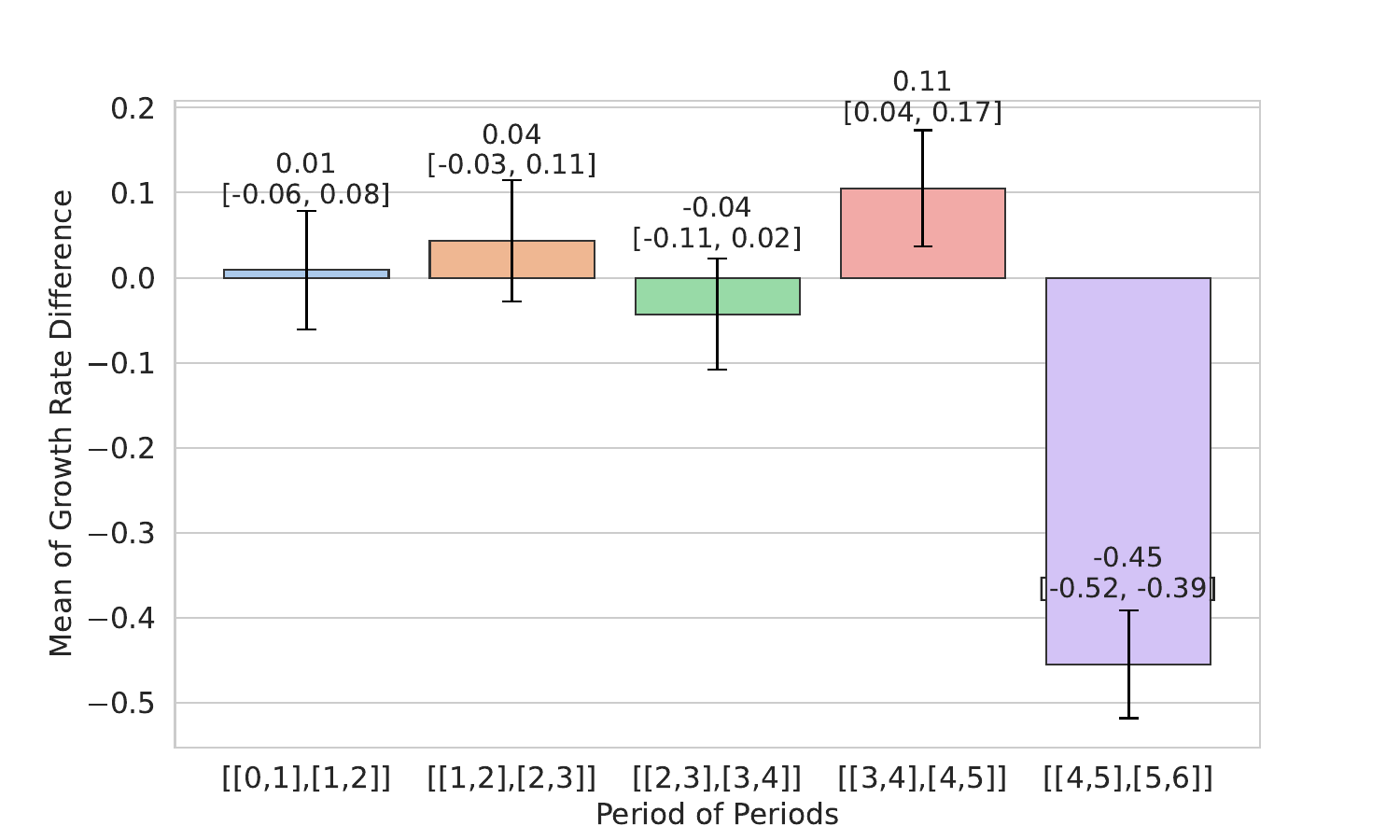}
     }\
        \caption{ Mean values and 99\% confidence interval of the ego network sizes in (a) and of the difference of their growth rate in (b).}
        \label{fig:egonet_size}
    \vspace{-10pt}
\end{figure}

We start our analysis by looking at the impact of reduced physical socialization on online ego networks. We will analyze the dimension of the active ego network size $|A_i^u|$ for all users $u$ during the period $I_i$ and their growth rate\footnote{With $A_i^u$ we indicate the active network (i.e., the set of active alters), while $|\cdot|$ indicates the cardinality of a set.}.
Figure~\ref{fig:egonet_size} (a) shows the average values and confidence intervals for the sizes of active ego networks $|A_i^u|$ across each period $I_i$. Two key observations emerge from the data. First, there is a significant increase in ego network sizes during the lockdown period ($I_5$), representing the largest growth across all periods examined. This is further highlighted by the fact that the confidence interval for $I_5$ does not overlap with those of other years. Second, there is a decrease in network size in the subsequent period $(I_6)$, which is the only observed reduction size in the dataset. These findings suggest that the expansion in ego network sizes during the lockdown period may be linked to users allocating more cognitive resources to online socialization, as most interactions were conducted online during that time. In contrast, the following year, as restrictive pandemic measures were relaxed, online socialization appeared to revert to pre-lockdown levels. Additionally, the data indicates a steady growth in ego network size in the years preceding the lockdown (i.e. periods $I_0, I_1, I_2, I_3, I_4$), likely driven by the increasing popularity of Twitter. However, the spike in network size during the lockdown phase (i.e., $I_5$) was particularly pronounced compared to previous trends.

To support our claim, we analyze the difference in the growth rate of the active ego network size across the time periods $(I_{i-1}, I_i, I_{i+1})$:
\begin{equation}
G_{[i+1,i, i-1]}^u(|A|) = G_{[i+1,i]}^u(|A|) - G_{[i,i-1]}^u(|A|) = \frac{|A|_{i+1}^u - |A|_{i}^u}{|A|_{i}^u} - \frac{|A|_{i}^u - |A|_{i-1}^u}{|A|_{i-1}^u} 
\end{equation}
A positive growth rate $G_{[i+1,i]}^u(|A|)$ indicates that the active ego network of user $u$ expands more rapidly during $I_{i+1}$ than during $I_i$ (Figure \ref{fig:egonet_size}b), while a negative growth rate indicates the opposite. To validate our findings, we conducted two t-tests, with the significance level set to 1\%, on these distributions, defining the following hypotheses:
\begin{itemize}
    \item $H_0^-$: The null hypothesis that the difference is non-positive. 
    \item $H_0^+$: The null hypothesis that the difference is non-negative. 
\end{itemize}
The results, shown in Table \ref{table:ttest_diff_sizes}, reveal that the first test rejected the non-positivity hypothesis ($H_0^-$) only for the triplet of periods $(I_3, I_4, I_5)$, with a p-value of $0.000$. This indicates that the growth rate of the active ego network between periods $I_4$ and $I_5$ is significantly higher compared to the growth rate between $I_3$ and $I_4$. In other words, \emph{the increase in the size of the active ego network is statistically significant only during the lockdown period} (i.e., $I_5$). Conversely, the second test rejects the non-negativity hypothesis ($H_0^+$) for the last triplet (i.e., $I_4, I_5, I_6$) with the same p-value, suggesting that \emph{the decrease in the size of the active ego network is statistically significant only in the post-lockdown period} (i.e., $I_6$).
For the earlier triplets of periods (i.e., $(I_0, I_1, I_2), (I_1, I_2, I_3), (I_2, I_3, I_4)$), there is no statistical evidence of any consistent increasing or decreasing trend. This aligns with Figure \ref{fig:egonet_size}(b), where the confidence interval and mean for the triplet of periods $(I_3, I_4, I_5)$ are significantly greater than zero, while for $(I_4, I_5, I_6)$, these values are significantly less than zero. For the earlier periods, the mean is close to zero, with the lower bound of the confidence interval negative and the upper bound positive.

\begin{table}[t!]
\caption{t-tests, with the significance level of $p$-values set to 1\%, of the difference of the growth rate of active ego network sizes.}\label{table:ttest_diff_sizes}
\setlength{\tabcolsep}{10pt}
\centering
\begin{tabular}{c ll ll}
\toprule
\multirow{2}{*}{periods} & \multicolumn{2}{c}{$H_0^-: \,\mathbb{E}[G_{[i+1,i, i-1]}^u(|A|)]\leq 0$} & \multicolumn{2}{c}{$H_0^+: \,\mathbb{E}[G_{[i+1,i, i-1]}^u(|A|)]\geq 0$}\\
\cmidrule(lr){2-3} \cmidrule(lr){4-5}
& outcome & $p$-value&outcome & $p$-value\\
\midrule
$(I_0,I_1,I_2)$ & ACCEPTED & $0.3696$ & ACCEPTED & $0.6304$ \\
$(I_1,I_2,I_3)$ & ACCEPTED & $0.0594$ & ACCEPTED & $0.9406$ \\
$(I_2,I_3,I_4)$ & ACCEPTED & $0.9538$ & ACCEPTED & $0.0462$ \\
$(I_3,I_4,I_5)$ & \textbf{REJECTED} & $0.0000$ & ACCEPTED & $1.0000$ \\
$(I_4,I_5,I_6)$ & ACCEPTED & $1.0000$ & \textbf{REJECTED} & $0.0000$ \\
\bottomrule
\end{tabular}
\vspace{-10pt}
\end{table}

Due to space constraints, we report here only the key message regarding the growth of ego network size. For a detailed analysis, we refer the interested reader to~\cite{cekini2024socialisolationdigitalconnection}, which highlights the following additional findings:
\begin{itemize}
    \item As ego networks expand, they develop additional circles, then return to their pre-lockdown structure once restrictions are lifted. The outermost circles experience the most growth, while the innermost circles remain relatively stable.
    \item Following lockdown, alters tend to shift toward inner circles, indicating a strengthening of relationships and increased intimacy during this period.
    \item After lockdown, egos gain a significant number of new alters and lose fewer than usual. However, once restrictions ease, many alters are dropped and fewer are added, suggesting that egos have reallocated their social cognitive resources, likely toward offline relationships.
\end{itemize}

\subsection{Signed Ego Networks Pre and Post Lockdown}
\label{sec:res_signed_ego_network}

While the analysis of structural variations in ego networks before and after lockdown (Section~\ref{sec:ego_network_analysis}) focuses on quantifying social interactions, signed ego networks add a novel qualitative dimension by capturing the positive or negative nature of relationships.
%
Positive links represent trust and affinity, while negative links indicate conflict or distrust. In this section, we analyze how the signs of social relationships vary before and after lockdown by focusing on the signed ego networks of regular and active users across each period $I_k$ for $k = 0, \dots, 6$ (i.e., the users $\bigcap_{i=0}^6 R_{I_i} \setminus \bigcup_{i=0}^6 O_{I_i}$ discussed in Section~\ref{sec:the_dataset}). The impact of reduced physical socialization on online ego networks id measured using the percentages of positive relationships ($P_i^u$) and negative relationships ($N_i^u$) for all users $u$ during the period $I_i$, as well as their growth rates.

\begin{figure}[t!]
     \centering
     \subfloat[Negative relationships]{
         \includegraphics[width=0.48\textwidth]{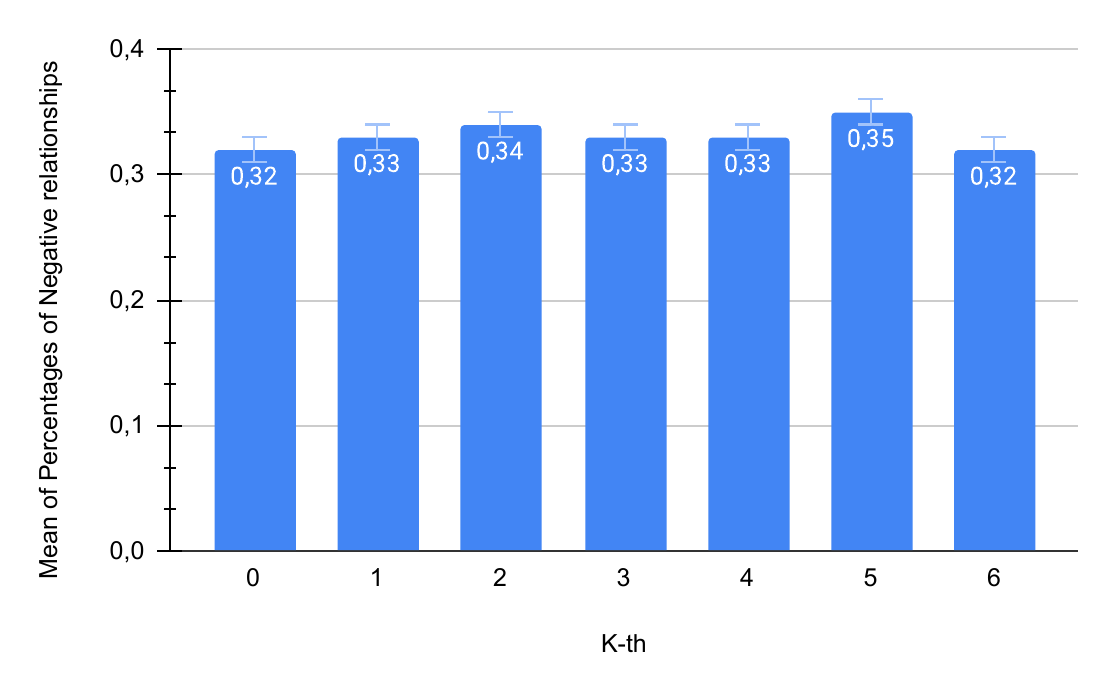}
     }\hfill
     \subfloat[Positive relationships]{
         \includegraphics[width=0.48\textwidth]{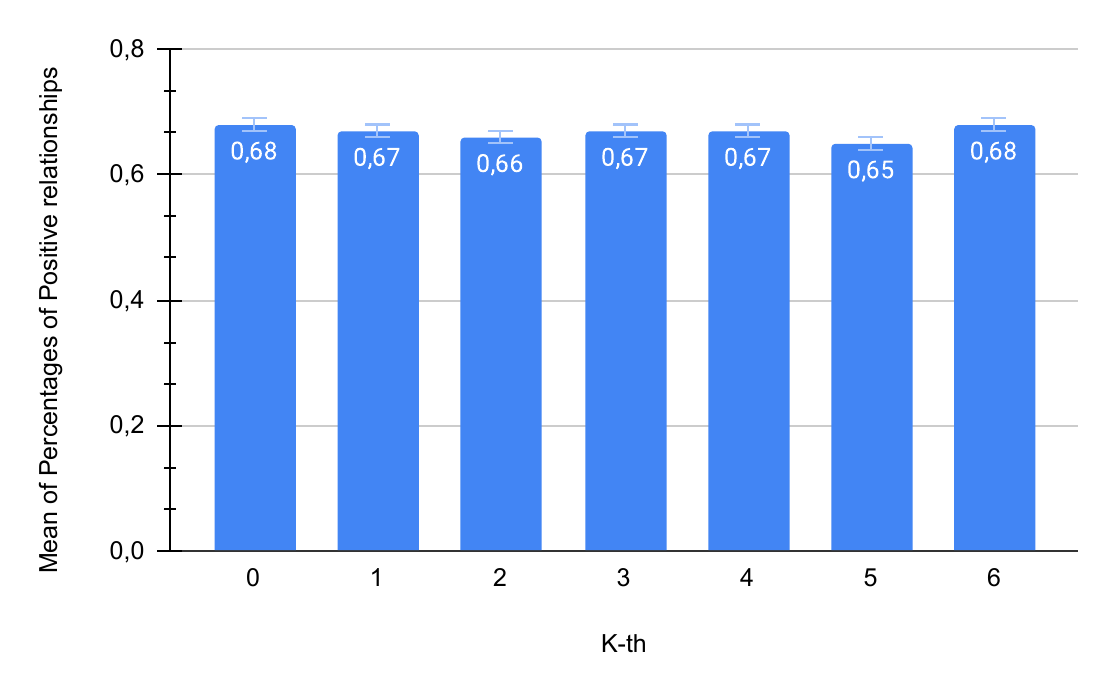}
     }\
        \caption{Mean values and 99\% confidence intervals for the percentage of negative relationships (a) and positive relationships (b).}
        \label{fig:percentage_negative_positive}
    \vspace{-10pt}
\end{figure}

Figure~\ref{fig:percentage_negative_positive} presents the mean and the 99\% confidence interval of the percentages of negative ($N_i^u$) and positive ($P_i^u$) relationships for each period $I_i$. Two key observations emerge from the data (supported by non-overlapping confidence intervals). First, during the lockdown, there is an increase in negative relationships, accompanied by a decrease in positive ones. Second, in the post-lockdown period ($I_6$), the opposite trend is observed: a reduction in negative relationships and an increase in positive ones. Excluding the initial period $I_0$, the percentages of positive and negative relationships remain relatively stable in the other periods. The rise in negative relationships during the lockdown can be attributed to the intensification of social and psychological distress caused by the restrictions~\citep{xiong2020impact}. Social isolation, economic uncertainty, and fear of the pandemic may have worsened social interactions, increasing conflicts and reducing positive interactions. In the post-lockdown period, the gradual lifting of restrictions favored the restoration of positive interactions, bringing them back to pre-lockdown levels.

\begin{figure}[t!]
     \centering
     \subfloat[Negative relationships]{
         \includegraphics[width=0.48\textwidth]{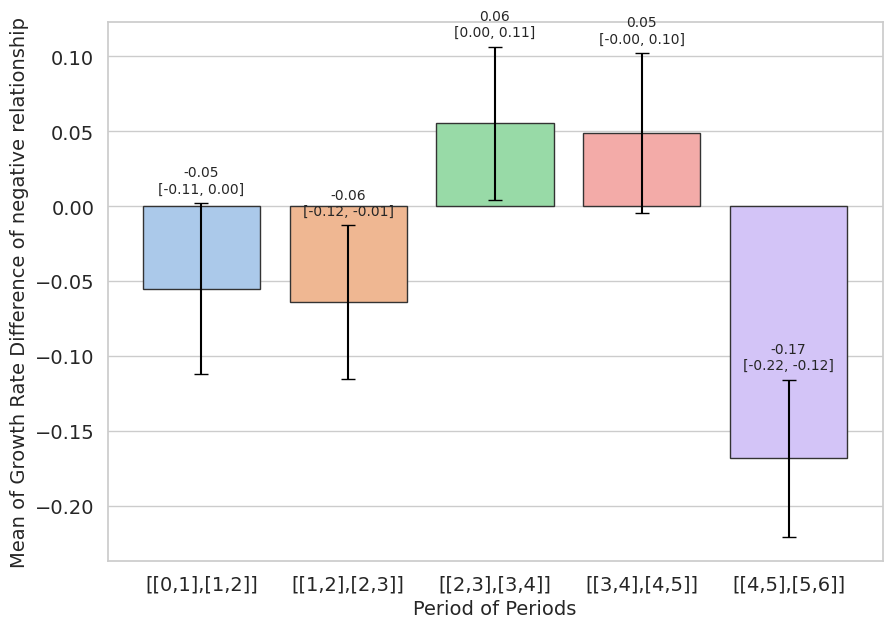}
     }\hfill
     \subfloat[Positive relationships]{
         \includegraphics[width=0.48\textwidth]{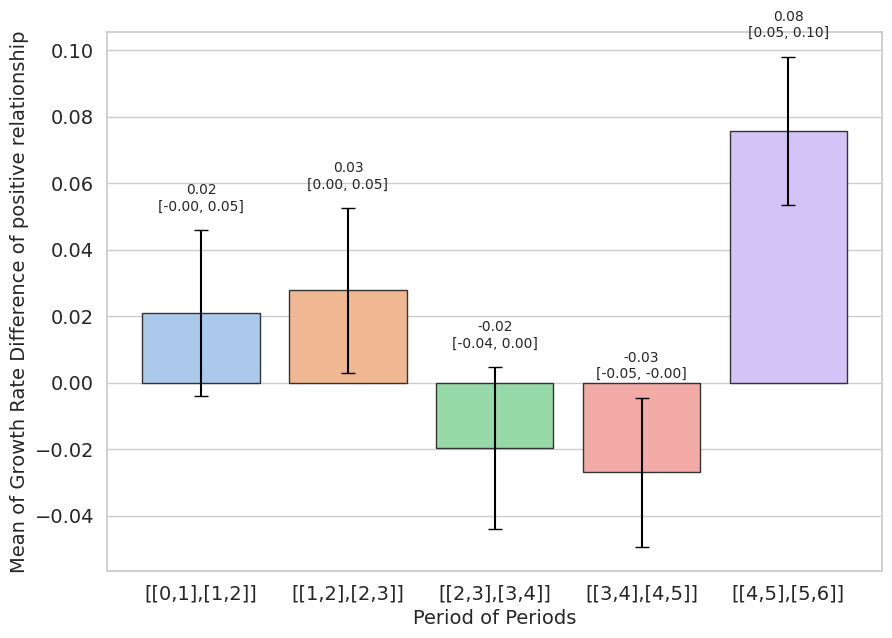}
     }\
        \caption{Mean values and 99\% confidence intervals for the growth rates of the difference in the percentage of negative relationships (a) and positive relationships (b).}
        \label{fig:growth_percentage_negative_positive}
        \vspace{-10pt}
\end{figure}

To support this claim, we analyzed the variation in the growth rates of negative and positive relationships across different periods $(I_{i-1}, I_i, I_{i+1})$, formally expressed as:
\begin{equation}
 G_{[i+1,i, i-1]}^u(N) = G_{[i+1,i]}^u(N) - G_{[i,i-1]}^u(N) = \frac{N_{i+1}^u - N_{i}^u}{N_{i}^u} - \frac{N_{i}^u - N_{i-1}^u}{N_{i-1}^u}
\end{equation}
\begin{equation}
 G_{[i+1,i, i-1]}^u(P) = G_{[i+1,i]}^u(P) - G_{[i,i-1]}^u(P) = \frac{P_{i+1}^u - P_{i}^u}{P_{i}^u} - \frac{P_{i}^u - P_{i-1}^u}{P_{i-1}^u}.
\end{equation}
A positive growth rate in $G_{[i+1,i]}^u(N)$ indicates that negative relationships are increasing faster between $I_{i}$ and $I_{i+1}$, whereas a negative growth rate suggests the opposite (Figure \ref{fig:growth_percentage_negative_positive} (a)). The same logic applies to positive relationships (Figure \ref{fig:growth_percentage_negative_positive} (b)). 
From Figure~\ref{fig:growth_percentage_negative_positive} there is strong evidence that negative relationships significantly decrease during the post-pandemic ($I_6$), while positive relationships significantly increase. The opposite happens during the lockdown ($I_5$). However, since the same trend can be observed in the period before, we cannot argue that this is a lockdown effect, at least with a visual analysis. 

To test the validity of our intuitions, we applied two t-tests (significance level 1\%) to the distribution of growth rate differences for relationship polarity, evaluating the hypotheses $H_0^-$ and $H_0^+$ previously defined.
For negative relationships (Table~\ref{table:ttest_diff_perc_negative}), the first test rejected the null hypothesis of non-positivity ($H_0^-$) for the triplet of periods $(I_3, I_4, I_5)$ with a $p$-value of $0.009$, indicating that the growth rate of negative relationships between $I_4$ and $I_5$ is significantly higher than between $I_3$ and $I_4$. Conversely, the second test rejected the null hypothesis of non-negativity ($H_0^+$) for the triplet of periods $(I_4, I_5, I_6)$ with a $p$-value of $0.0 \times 10^{-4}$, suggesting a statistically significant decrease in negative relationships in the post-lockdown phase (i.e. from $I_5$ to $I_6$) compared to the lockdown phase (i.e. from $I_4$ to $I_5$).
On the other hand, Table~\ref{table:ttest_diff_perc_positive} shows the results for positive relationships, where the second t-test rejects the null hypothesis ($H_0^+$) for the triplet periods $(I_3, I_4, I_5)$ with a $p$-value of $0.01$, indicating that \emph{the decline in positive relationships between $I_4$ and $I_5$ is statistically significant}. Conversely, the first test rejects the null hypothesis of non-positivity ($H_0^-$) for the final triplet of period $(I_4, I_5, I_6)$ with a $p$-value of $0.0 \times 10^{-4}$, suggesting that \emph{the increase in positive relationships in the post-lockdown period is statistically significant}.
For the earlier periods, excluding $(I_1, I_2, I_3)$, there is no statistical evidence to suggest significant trends in either negative or positive relationships.

\begin{table}[!t]
\caption{t-tests, with the significance level of $p$-values set to 1\%, of the difference of the growth rate of the consecutive percentages of Negative relationship.}\label{table:ttest_diff_perc_negative}
\setlength{\tabcolsep}{10pt}
\centering
\begin{tabular}{c ll ll}\\ 
\toprule
\multirow{2}{*}{periods} & \multicolumn{2}{c}{$H_0^-: \mathbb{E}[G_{[i+1,i, i-1]}^u(N)] \le 0$} & \multicolumn{2}{c}{$H_0^+: \mathbb{E}[G_{[i+1,i, i-1]}^u(N)] \ge 0$}\\
\cmidrule(lr){2-3} \cmidrule(lr){4-5}
& outcome & $p$-value&outcome & $p$-value\\
\midrule
$(I_0,I_1,I_2)$ & ACCEPTED & $0.99$& \textbf{ACCEPTED} & $0.01$\\
$(I_1,I_2,I_3)$ & ACCEPTED & $0.48$& ACCEPTED & $0.52$\\
$(I_2,I_3,I_4)$ & ACCEPTED & $0.35$ & ACCEPTED & $0.65$ \\
$(I_3,I_4,I_5)$ & \textbf{REJECTED}& $0.00$& ACCEPTED & $1.00$\\
$(I_4,I_5,I_6)$ & ACCEPTED& $1.00$& \textbf{REJECTED} & $0.00$\\
\end{tabular}
\vspace{-20pt}
\end{table}

\begin{table}[!t]
\caption{t-tests, with the significance level of $p$-values set to 1\%, of the difference of the growth rate of the consecutive percentages of Positive relationship.}\label{table:ttest_diff_perc_positive}
\setlength{\tabcolsep}{10pt}
\centering
\begin{tabular}{c ll ll}\\ 
\toprule
\multirow{2}{*}{periods} & \multicolumn{2}{c}{$H_0^-: \mathbb{E}[G_{[i+1,i, i-1]}^u(P)] \le 0$} & \multicolumn{2}{c}{$H_0^+: \mathbb{E}[G_{[i+1,i, i-1]}^u(P)] \ge 0$}\\
\cmidrule(lr){2-3} \cmidrule(lr){4-5}
& outcome & $p$-value&outcome & $p$-value\\
\midrule
$(I_0,I_1,I_2)$ & \textbf{REJECTED}& $0.00$& ACCEPTED & $1.00$\\
$(I_1,I_2,I_3)$ & ACCEPTED & $0.05$& ACCEPTED & $0.95$\\
$(I_2,I_3,I_4)$ & ACCEPTED & $0.93$& ACCEPTED & $0.07$\\
$(I_3,I_4,I_5)$ & ACCEPTED & $0.99$& \textbf{REJECTED} & $0.01$\\
$(I_4,I_5,I_6)$ & \textbf{REJECTED}& $0.00$& ACCEPTED & $1.00$ \\
\end{tabular}
\vspace{-20pt}
\end{table}

\subsection{Topic Evolution Pre and Post Lockdown}
\label{sec:semantic_topic_evolution}

After analyzing network structure (Section~\ref{sec:ego_network_analysis}) and relationship polarity (Section~\ref{sec:res_signed_ego_network}), we now examine the content of user interactions. We investigate if the behavioral changes seen during lockdown – such as larger networks and more negative links – also correspond to changes in what users discussed online.
For each social interaction (with the exception of retweets), the topics are extracted as discussed in Section~\ref{sec:BERTopic}.

To quantify the thematic diversity in user ego networks, we compute the number of unique topics \( |T_j^{I_i}| \) for each user \( j \) in each period \( I_i \), excluding retweets and topics identified as outliers by HDBSCAN (as mentioned in Section~\ref{sec:BERTopic}). This metric captures the variety of topics discussed by each user. Our hypothesis is that the lockdown, by reshaping online social dynamics, also influenced topic diversity. Specifically, we test whether the variation in unique topics follows the pattern observed in structural metrics: a peak during the main lockdown period (\( I_5 \)), followed by a return to previous levels (\( I_6 \)).

\begin{figure}[t!] 
    \centering 

    \begin{subfigure}[b]{0.48\textwidth} 
        \centering
        \includegraphics[width=\linewidth]{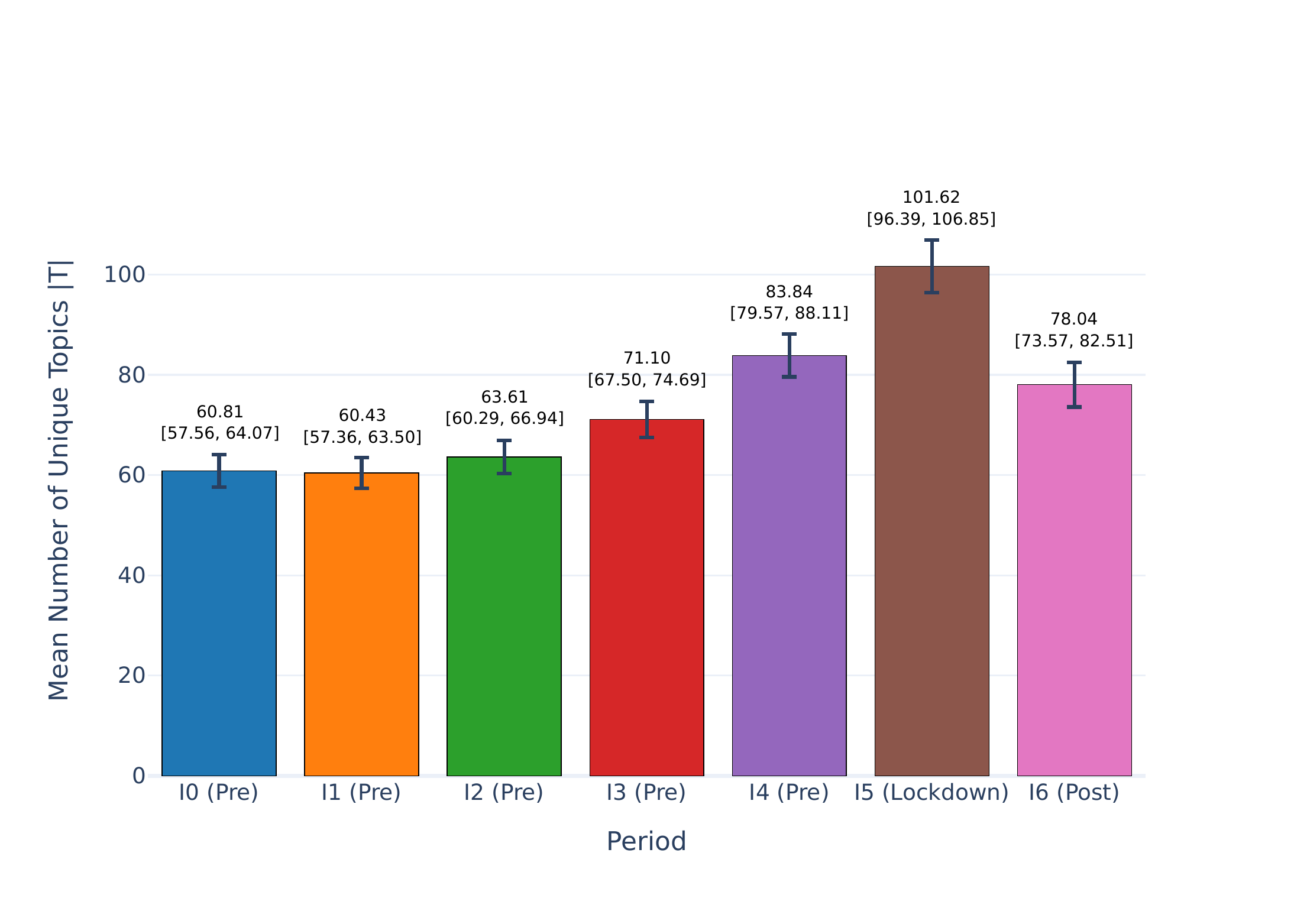} 
        \caption{Mean number of unique topics per user ($\mathbb{E}[|T|]$) across periods ($I_0$-$I_6$).} 
        \label{fig:mean_topics} 
    \end{subfigure}
    \hfill 
    \begin{subfigure}[b]{0.48\textwidth} 
        \centering
        \includegraphics[width=\linewidth]{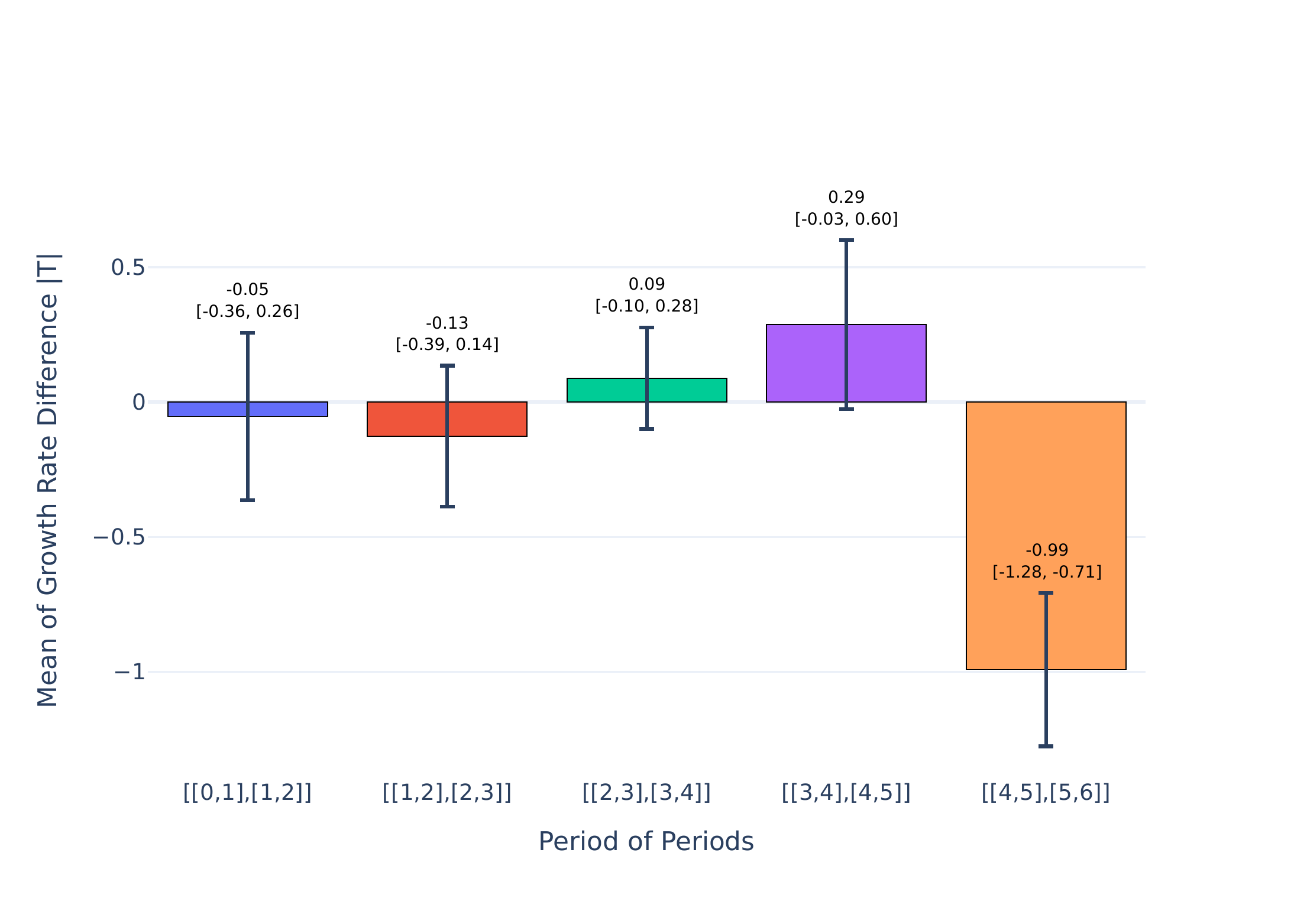} 
        \caption{Mean difference in consecutive annual growth rates ($D_k = G_k(|T|) - G_{k-1}(|T|)$).} 
        \label{fig:diff_growth} 
    \end{subfigure}
    \caption{Average topic diversity in ego networks across periods (outliers are not counted). (a) Mean unique topics ($\mathbb{E}[|T|]$) per period. (b) Mean difference between consecutive annual growth rates ($D_k$). Error bars show the 95\% CI.}
    \label{fig:combined_topic_analysis} 
\end{figure}

Figure~\ref{fig:combined_topic_analysis} (a) shows the average number of unique topics ($|T_{I_i}^j|$) per user for each period $I_i$. The analysis reveals two main points:
\begin{itemize}
    \item \textbf{Significant Increase During Lockdown:} The number of valid unique topics rises significantly during the lockdown period (\( I_5 \)), representing the largest increase among all examined periods. This finding is further supported by the fact that the confidence interval for \( I_5 \) does not overlap with those of other periods.
    \item \textbf{Post-Lockdown Decline:} In the subsequent period, \( I_6 \), there is a noticeable decrease in the number of valid unique topics - the only decline recorded in the entire dataset. This suggests that the expansion of unique topics during lockdown may have been driven by increased cognitive engagement in online social interactions, a trend that reversed post-lockdown.
\end{itemize}


\noindent
Figure~\ref{fig:combined_topic_analysis} (a) indicates a steady growth in the number of unique topics during the periods preceding the lockdown (i.e., \( I_2, I_3, \) and \( I_4 \)); however, the peak observed in \( I_5 \) is significantly more pronounced than previous trends. To further validate these findings, we examined the change in the growth rate of the number of unique topics between consecutive periods. Specifically, we analyzed the difference in the growth rate of unique topics between periods \((I_{i-1}, I_i, I_{i+1})\) using the following formulation:
\begin{equation}
G_u [i+1,i,i-1](|T|) = G_u [i+1,i](|T|) - G_u [i,i-1](|T|) = \frac{|T_{I_{i+1}}^u| - |T_{I_i}^u|}{|T_{I_i}^u|} - \frac{|T_{I_i}^u| - |T_{I_{i-1}}^u|}{|T_{I_{i-1}}^u|}
\end{equation}
A positive value of \( G_u [i+1,i] \) indicates that the number of unique topics for user \( u \) increases more rapidly in period \( I_{i+1} \) than in \( I_i \), as highlighted in Figure~\ref{fig:combined_topic_analysis} (b). Conversely, a negative value suggests the opposite.
To validate these results, we applied two t-tests (significance level 5\%) to the distribution of growth rate differences for unique topics, evaluating the hypotheses $H_0^-$ and $H_0^+$ previously defined.
The results are summarized in Table~\ref{table:ttest_diff_sizes_topics}. For the triplet \((I_3, I_4, I_5)\), rejection of \( H^-_0 \) (with a p-value of \( 0.0 \times 10^{-4} \)) confirms that the growth rate of unique topics between \( I_4 \) and \( I_5 \) is significantly higher than that observed between \( I_3 \) and \( I_4 \). Figure~\ref{fig:combined_topic_analysis} (b) supports this finding, displaying a mean above zero, indicative of a \emph{marked increase during lockdown}. For the triplet \((I_4, I_5, I_6)\), rejection of \( H^+_0 \) suggests that the \emph{decrease in the number of unique topics in the post-lockdown period is statistically significant}. Figure~\ref{fig:combined_topic_analysis} (b) confirms this result by showing a confidence interval and mean below zero, indicative of a significant reduction in the growth rate during this period.
For the earlier triplets of periods (i.e., $(I_0, I_1, I_2), (I_1, I_2, I_3), (I_2, I_3, I_4)$), there is no statistical evidence of any consistent increasing or decreasing trend. This is consistent with Figure~\ref{fig:combined_topic_analysis} (b), where the confidence interval and mean for the triplet of periods $(I_3, I_4, I_5)$ are significantly greater than zero, while for $(I_4, I_5, I_6)$, these values are significantly less than zero. For the earlier periods, the mean is close to zero, with the lower bound of the confidence interval negative and the upper bound positive.


\begin{table}[t!]
\caption{t-tests, with the significance level of $p$-values set to 5\%, of the difference of the growth rate of the number of unique topics.}\label{table:ttest_diff_sizes_topics}
\setlength{\tabcolsep}{10pt}
\centering
\begin{tabular}{c ll ll}
\toprule
\multirow{2}{*}{periods} & \multicolumn{2}{c}{$H_0^-: \,\mathbb{E}[G_u [i+1,i,i-1](|T|) ]\leq 0$} & \multicolumn{2}{c}{$H_0^+: \,\mathbb{E}[G_u [i+1,i,i-1](|T|) ]\geq 0$}\\
\cmidrule(lr){2-3} \cmidrule(lr){4-5}
& outcome & $p$-value&outcome & $p$-value\\
\midrule
$(I_0,I_1,I_2)$ & ACCEPTED & $0.633$& ACCEPTED & $0.367$\\
$(I_1,I_2,I_3)$ & ACCEPTED & $0.8285$& ACCEPTED & $0.1715$\\
$(I_2,I_3,I_4)$ & ACCEPTED & $0.1782$& ACCEPTED & $0.8218$\\
$(I_3,I_4,I_5)$ & \textbf{REJECTED} & $0.0364$& ACCEPTED & $0.9636$\\
$(I_4,I_5,I_6)$ & ACCEPTED & $1.0000$ & \textbf{REJECTED} & $0.0000$ \\
\bottomrule
\end{tabular}
\end{table}

\section{Conclusions}
This study explores how COVID-19 restrictions influenced users’ online social behavior by examining multiple facets of their interactions. We concentrated on users’ ego networks, assessing their structural properties, the sentiment of their relationships (through signed networks), and the thematic content of their communication (via semantic networks), each offering a unique lens into online social cognitive processes. The analysis is based on a dataset of over 1,000 Twitter users, whose timelines span a seven-year period—covering five years prior to the COVID-19 lockdown and two years following it.

Our findings reveal substantial changes across all examined dimensions during the lockdown. From a structural standpoint, ego networks grew in size, with more alters and additional circles—particularly in the outer layers—while existing alters tended to shift toward more central circles, reflecting deeper ties. In terms of relationship sentiment, we detected a significant rise in negative interactions within the ego networks. Finally, the semantic analysis showed that users began engaging with a broader array of topics, indicating increased thematic diversity in their communications. The overall takeaway is that Twitter users increased their cognitive engagement online, both in terms of social interactions and thematic diversity. As expected, the negativity of social exchanges also increased after lockdown, reflecting the difficult period users were living in.
Interstingly, these changes appeared to be largely temporary adaptations to the unique context of the lockdown. Once restrictions were relaxed, structural metrics, the balance of relationship polarity, and thematic diversity tended to return to their pre-pandemic levels. These temporary changes seem driven by different factors. Limited offline contact likely allowed users to allocate more cognitive resources to their online engagements, thereby expanding their networks and exploring a broader range of topics. At the same time, the stress of the pandemic~\citep{xiong2020impact} period probably contributed to the increase in negative online relationships. As the lockdown ended, users likely returned to offline activities and experienced less stress, explaining why online networks reverted to their pre-pandemic state across all observed metrics.

\bmhead{Funding}

This work was partially supported by SoBigData.it. SoBigData.it receives funding from European Union – NextGenerationEU – National Recovery and Resilience Plan (Piano Nazionale di Ripresa e Resilienza, PNRR) – Project: “SoBigData.it – Strengthening the Italian RI for Social Mining and Big Data Analytics” – Prot. IR0000013 – Avviso n. 3264 del 28/12/2021. 
This work is also supported by the European Union under the scheme HORIZON-INFRA-2021-DEV-02-01 – Preparatory phase of new ESFRI research infrastructure projects, Grant Agreement n.101079043, “SoBigData RI PPP: SoBigData RI Preparatory Phase Project”. 
The work of K. Cekini, M. Conti and A. Passarella is partly supported by PNRR - M4C2 - Investimento 1.3, Partenariato Esteso PE00000013 - ``FAIR - Future Artificial Intelligence Research'' - Spoke 1 "Human-centered AI", funded by the European Commission under the NextGeneration EU programme. 
The work of E. Biondi is partly supported by project SERICS (PE00000014) under the MUR National Recovery and Resilience Plan funded by the European Union - NextGenerationEU.
C. Boldrini was also supported by PNRR - M4C2 - Investimento 1.4, Centro Nazionale CN00000013 - "ICSC - National Centre for HPC, Big Data and Quantum Computing" - Spoke 6, funded by the European Commission under the NextGeneration EU programme.

%
%

\bigskip

\begin{appendices}






\section{Supplementary Information}
\label{sec:appendix_details}

\subsection{Hyperparameter Optimization for BERTopic}
\label{subsec:appendix_hyperparam} 

As mentioned in the body of the paper, the performance of the BERTopic pipeline depends on the hyperparameter configuration. To identify the optimal configuration for clustering, a \textit{grid search} was performed by testing different combinations of key hyperparameters.
The parameters considered in the optimization were as follows:
\begin{itemize}
    \item \textbf{\texttt{n\_components} (UMAP):} Specifies the number of dimensions into which UMAP projects the data. Too low a dimensionality might lead to the loss of relevant information, while too high a dimensionality could diminish the computational advantages of dimensionality reduction.\\
    \textit{Explored Values:} $\{2, 3, 5, 10, 50, 100\}$.

    \item \textbf{\texttt{n\_neighbors} (UMAP):} Defines the number of neighbors considered by UMAP to compute the local structure of the data. This parameter balances the preservation of local versus global structure: low values emphasize local structure, while higher values tend to preserve more global relationships.\\
    \textit{Explored Values:} $\{5, 10, 15, 20, 50\}$.

    \item \textbf{\texttt{min\_cluster\_size} (HDBSCAN):} Determines the minimum size a group of points must have to be considered a cluster. It influences the granularity of the clustering and the model's ability to identify smaller, potentially significant clusters.\\
    \textit{Explored Values:} $\{20, 40, 80, 120, 140, 160, 200, 220\}$.

    \item \textbf{\texttt{cluster\_selection\_method} (HDBSCAN):} Specifies the criterion used by HDBSCAN to select the final clusters from the hierarchy. Both the \texttt{eom} (Excess of Mass) and \texttt{leaf} methods were tested. The \texttt{eom} method often identifies a few large clusters alongside many small ones, whereas the \texttt{leaf} method prioritizes cluster stability, typically producing more homogeneous and detailed clusters. Experimental results showed greater consistency and better cluster granularity with the \texttt{'leaf'} option, which was therefore chosen for the final configuration, as it enhances the model's ability to distinguish topics in noisy and multilingual datasets like ours.
\end{itemize}
In summary, the grid search was conducted by exploring the space defined by the Cartesian product of the values listed above:
\[ \{2,3,5,10,50,100\} \times \{5,10,15,20,50\} \times \{20,40,80,120,140,160,200,220\} \times \{\texttt{'leaf'}\} \]

The top 10 hyperparameter configurations identified using the sampled DBCV score during the grid search are reported in Table~\ref{tab:hyperpar_dbcv}.

\begin{table}[h!] 
\centering
\renewcommand{\arraystretch}{1.3} 
\begin{tabular}{ccccc}
\toprule
\textbf{n\_components} & \textbf{n\_neighbors} & \textbf{min\_cluster\_size} & \textbf{Outliers (\%)} & \textbf{DBCV Samp. Score} \\
\midrule
2  & 10 & 20  & 72.66 & 0.11766 \\
10 & 15 & 160 & 80.21 & 0.11023 \\
10 & 15 & 140 & 80.74 & 0.11012 \\
10 & 15 & 120 & 80.93 & 0.10924 \\
5  & 10 & 160 & 80.92 & 0.10780 \\
10 & 15 & 80  & 81.17 & 0.10629 \\
5  & 10 & 140 & 80.83 & 0.10604 \\
5  & 10 & 120 & 80.63 & 0.10493 \\
5  & 15 & 120 & 80.60 & 0.10415 \\
5  & 15 & 160 & 80.54 & 0.10365 \\
\bottomrule
\end{tabular}
\caption{Top 10 configurations based on sample DBCV score.}
\label{tab:hyperpar_dbcv} 
\end{table}

\subsection{Additional Dataset Characteristics}
\label{app:dataset_context} 

\begin{figure}[h!]
     \centering
     \includegraphics[width=1\textwidth]{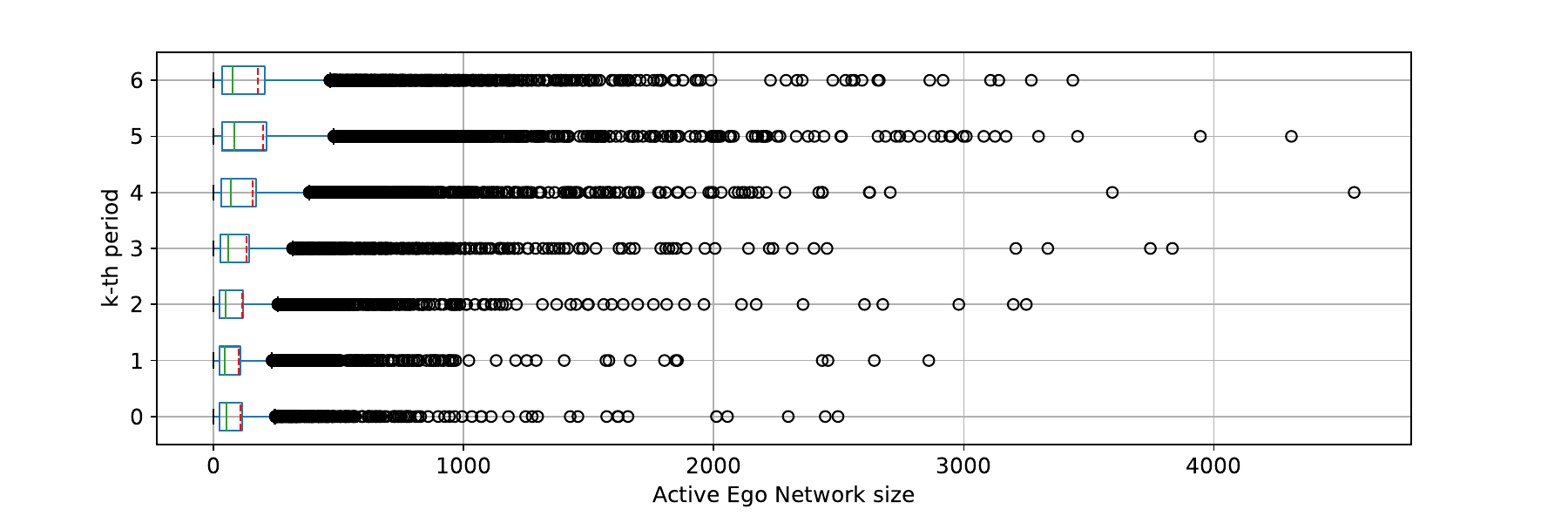}
    \caption{Distribution of active ego networks sizes}
    \label{fig:egonet_size_all} \vspace{-20pt}
\end{figure}

This section provides supplementary details regarding the dataset downloaded and the user filtered for the study. In table~\ref{tab:users_filtered} we detail the total number of users from the dataset and the number obtained at each step of the filtering applied (bot removal, regular and active users selection and outlier removal), as described in Section~\ref{sec:the_dataset}. Moreover, in Figure~\ref{fig:egonet_size_all} there is the distribution of active ego network sizes, where are highlighted the outliers $O_{I_i}$ in each time period $I_i$.

\begin{table}[h!]
\centering
\renewcommand{\arraystretch}{1.5} 
\begin{tabular}{@{}lr@{}}
\toprule
\textbf{User Category} & \textbf{Number of users} \\ \midrule
All Users& 53,837 \\ \hline
Human users & 10,547 \\ \hline
$\bigcap_{i=0}^6 R_{I_i}$ & 1,627 \\ \hline
$\bigcap_{i=0}^6 R_{I_i} \setminus \bigcup_{i=0}^6 O_{I_i}$ & 1,286 \\ \bottomrule
\end{tabular}
\caption{Summary of applied filters}
\label{tab:users_filtered}
\vspace{-20pt}
\end{table}

\end{appendices}


\bibliography{sn-bibliography}

\end{document}